\documentclass[12pt,a4paper]{article}
\usepackage{epsfig}
\usepackage{amsfonts}
\usepackage{amssymb}

\begin{document}
\begin{center}

\vspace{1cm}

{\bf \large ON THE AMPLITUDES IN $\mathcal{N}=(1,1)$ $D=6$ SYM} \vspace{1.5cm}

{\bf \large L. V. Bork$^{1}$, D. I. Kazakov$^{1,2,3}$, D. E. Vlasenko$^{2,3}$}\vspace{0.5cm}

{\it $^1$Institute for Theoretical and Experimental Physics, Moscow, Russia\\
$^2$Bogoliubov Laboratory of Theoretical Physics, Joint
Institute for Nuclear Research, Dubna, Russia.\\
$^3$Moscow Institute of Physics and Technology, Dolgoprudny, Russia}\vspace{0.5cm}

\abstract{We consider the on-shell amplitudes in $\mathcal{N}=(1,1)$ SYM
in $D=6$  dimensions within the spinor helicity and on-shell superspace formalism.
This leads to an effective and straightforward technique reducing the calculation to a set
of scalar master integrals.
As an example, the simplest four point amplitude  is calculated in one and two loops
in the planar limit. All answers are UV and IR finite and expressed in terms of logs and
Polylogs of transcendentality level 2 at one loop, and 4 and 3 at two loops.
The all loop asymptotical limit at high energy
is obtained which exhibits the Regge type behaviour. The intercept is calculated in the planar limit
and is equal to $\alpha(t)=1+\sqrt{\frac{g_{YM}^2 N_c|t|}{32\pi^3}}$.}
\end{center}


\newpage

\tableofcontents{}\vspace{0.5cm}

\section{Introduction}\label{Introduction_1}
In the last decade tremendous progress has been achieved in
understanding the structure of the $S$-matrix of four dimensional
gauge theories  (for review see, for example,
\cite{Reviews_Ampl_General}). The most impressive results have been
obtained in the theories with extended supersymmetry, the
$\mathcal{N}=4$ SYM is one of the important examples.

The $\mathcal{N}=4$ SYM in addition to ordinary (super)conformal
invariance has a new type of symmetry, dual (super)conformal
invariance, i.e., the conformal symmetry in momentum space. Taking
together, the algebras of these symmetries can be fused into an
infinite dimensional Yangian algebra which in principle should
completely define the $S$-matrix of the theory
\cite{BeisertYangianRev}. Also, the $\mathcal{N}=4$ SYM possesses
the unexpected dualities between the amplitudes and the Wilson
loops, the amplitudes and the correlation functions and presumably
between the form factors and the Wilson loops (see, for example,
\cite{Reviews_N=4SYMProperties} for review). In addition to the
above mentioned properties, the recent results suggest that in the
description of the structure of the $S$-matrix of the
$\mathcal{N}=4$ SYM  the language of  algebraic geometry (the motive
theory) might be useful \cite{Arcani-Hamed Grassmannians}. All
these intriguing properties of the $\mathcal{N}=4$ SYM are
intimately linked together and are not fully understood at the
current moment.

It should be noted that the above mentioned  results are almost  impossible to obtain using
the standard textbook computational methods. The new technique is intensively used:  the spinor helicity and momentum twistors formalisms,
different sets of recurrence relations for the tree level amplitudes,
the unitarity  based methods for loop amplitudes and different
realizations of the on-shell superspace technique for
theories with supersymmetry \cite{Reviews_Ampl_General}.

It is interesting to note that the spinor helicity formalism and  the unitarity based methods
can be generalised to space-time dimension greater than $D=4$
\cite{SpinorHelisityForm_D=10Dimentions, SpinorHelisityForm_GeneralDimentions_Boels}. So the
gauge theories in extra dimensions can  also be studied by these methods. For
example, the  spinor helicity formalism was suggested for $D=6$ in \cite{DonaldOConnel_AmplInD=6},
so the $S$-matrix in  $D=6$ gauge theory can be calculated like  its
$D=4$ counterpart.

In particular, the $D=6$ gauge theories with maximal supersymmetry,
namely, (1,1) and (2,0) supersymmetries, are of special interest
\cite{D=5SYM_and_(20)Theory,(20)Neil,AmplitudesAndM5Branes}. At the
tree level the amplitudes of the (1,1) SYM theory can be interpreted as
amplitudes of the $D=4$ $\mathcal{N}=4$ SYM theory with the Higgs
regulator \cite{DonaldOConnel_AmplInD=6, ZBern_GenUnit_D=6Helicity}.
Then, they can be used in the unitarity based computations of the
loop amplitudes in the $D=4$ $\mathcal{N}=4$ massive SYM. They  may
also be useful in QCD computations of rational terms of the one loop
amplitudes, and one can encounter other $D=6$ objects as parts of
the QCD multiloop computations \cite{D=6 QCD}. In addition, the (1,1) and (2,0)
theories can be considered as a special low energy limit (the effective
actions on the 5-branes) of the string/M theory. After additional
compactifications on two-torus both the theories reduce to the $D=4$
$\mathcal{N}=4$ SYM. One may wonder if the origin of dual
(super)conformal symmetry and other "miracles" of the
$\mathcal{N}=4$ SYM lies in the properties of the string theory in 6
dimensions \cite{Arcani-Hamed Grassmannians}. For these reasons it would be interesting to study the
$S$-matrix of the (1,1) and (2,0) $D=6$ gauge theories.

In this article we focus on the four point amplitudes in the (1,1)
$D=6$ SYM theory. As was explained above, we use the spinor helicity
formalism suggested in \cite{DonaldOConnel_AmplInD=6} and the
on-shell momentum superspace formalism proposed in
\cite{Sigel_D=6Formalism}. The tree level 3, 4 and 5 point
amplitudes were obtained in \cite{DonaldOConnel_AmplInD=6}.  In
addition to it, the  4- and 5-point amplitudes were studied in
\cite{Sigel_D=6Formalism,Brandhuber_1loopD=6} at the one loop level.
The symmetry properties of the amplitudes were discussed in
\cite{D6_DualConformal_Invariance}. Our main interest is the
structure of the four point amplitude at the multiloop level and
its comparison with the $D=4$ $\mathcal{N}=4$ SYM 4-point amplitude.

The article is organised as follows. In section 2, we discuss the
spinor helicity formalism and the general structure of the on-shell momentum superspace
in the $D=6$ (1,1) SYM. In section 3, we compute the  4-point amplitude in one and two loops in
terms of scalar integrals by means of the iterated unitarity cuts. In section 4, we
compute the scalar integrals and discuss the structure of the
corresponding amplitude. We also investigate its asymptotic Regge
behaviour of the 4-point amplitude in all loops  and obtain the Regge asymptotics.
In appendices, we  discuss the computation of the $D=6$ double box
integral by means of the MB representation and give the derivation of the all-loop Regge asymptotics  from the ladder diagrams.

\section{The spinor helicity formalism and  on-shell superspace in six dimensions}\label{D=6 on-shell superspace_2}
This section is based mostly on the papers \cite{DonaldOConnel_AmplInD=6} and
\cite{Sigel_D=6Formalism}.
We review first the $D=6$ spinor helicity formalism used throughout the calculations.

Consider the massless
$D=6$ vector $p^{\mu}$, $p^2=0$, $\mu=1,\ldots,6$ which transforms under the vector
representation of the $D=6$ Lorentz group $SO(5,1)$. Using the $D=6$ antisymmetric Pauli matrices
$(\sigma^{\mu})_{AB}$ and $(\overline{\sigma}_{\mu})^{AB}$, where the indices $A,B=1,\ldots,4$  transform under the fundamental
representation of the $Spin(SO(5,1))\simeq SU(4)^{*}$ (which is the covering group for $SO(5,1)$)\footnote{$SU(4)^{*}$ is a none compact generalization of $SU(4)$. One also has $Spin(SO(2,1))\simeq SL(2,\mathbb{R})$,$~Spin(SO(3,1))\simeq SL(2,\mathbb{C})$,$~Spin(SO(5,1))\simeq SL(2,\mathbb{Q})$, where $\mathbb{R}$,$\mathbb{C}$,$\mathbb{Q}$ are sets of real and complex
numbers and quaternions, respectively.} we can rewrite
$p^{\mu}$ by analogy with the $D=4$ case as:
\begin{eqnarray}
  p^{AB}=p^{\mu}(\overline{\sigma}_{\mu})^{AB},
\end{eqnarray}
or
\begin{eqnarray}
  p_{AB}=p_{\mu}(\sigma^{\mu})_{AB}.
\end{eqnarray}
Note that one can lower and rise the indices $A,B$ using absolutely
antisymmetric objects $\epsilon^{ABCD}$ and $\epsilon_{ABCD}$
associated with $SU(4)^{*}$:
\begin{eqnarray}
  (\sigma^{\mu})_{AB}=\frac{1}{2}\epsilon_{ABCD}(\overline{\sigma}_{\mu})^{CD}.
\end{eqnarray}
The condition $p^2=0$ in terms of the matrix $p^{AB}$ is equivalent to $\det(p)=0$,
so one can write $p^{AB}$ as a product of two commuting $SU(4)^{*}$ spinors:
\begin{eqnarray}
  p^{AB}=\lambda^{Aa}\lambda_{a}^B, \ \ \  p_{AB}=\tilde{\lambda}_{A}^{\dot{a}}\tilde{\lambda}_{B\dot{a}}.
\end{eqnarray}
Note that the spinors $\lambda^{Aa}$ and
$\tilde{\lambda}_{A}^{\dot{a}}$, in contrast to the $D=4$ case, in
addition to the covering group index $A$ also carry the little group
$SO(4) \simeq SU(2)\times SU(2)$ indices $a=1,2$ and
$\dot{a}=\dot{1},\dot{2}$. The little group for $D$ dimensions is
$SO(D-2)$, so  for $D=4$ it is just $SO(2) \simeq U(1)$ and the
action of the $D=4$ little group on spinors is just the
multiplication by a complex number $z$, $|z|=1$.  In the $D=6$ case
the action of the little group is no longer trivial and helicity is
no longer conserved in contrast to the $D=4$ case. Note also that
one cannot rise and lower the $SU(4)^{*}$ $A,B,\ldots$ indices for
spinors but  the little group indices $a$ and $\dot{a}$ using the
antisymmetric objects $\epsilon_{ab}$ and
$\epsilon_{\dot{a}\dot{b}}$ associated with the $SU(2)$ groups. Note
also that there are no any constraints on the spinors $\lambda^{Aa}$ and
$\tilde{\lambda}_{A}^{\dot{a}}$ \cite{Sigel_D=6Formalism} as in the $D=4$ case.

The Lorentz invariant products of spinors then can be given by:
\begin{eqnarray}
  \lambda(i)^{Aa}\tilde{\lambda}(j)_{A}^{\dot{a}}\doteq
  \langle i_a|j_{\dot{a}}]=[j_{\dot{a}}|i_a\rangle,
\end{eqnarray}
where $i$ and $j$ are the labels of external momenta $p^{\mu}_i$ and $p^{\mu}_j$
associated with the spinors $\lambda(i)^{Aa}$ and $\tilde{\lambda}(j)_{A}^{\dot{a}}$.
In addition, one has two Lorentz invariant  combinations of spinors:
\begin{eqnarray}
  \epsilon_{ABCD}\lambda(1)^{Aa}\lambda(2)^{Bb}\lambda(3)^{Cc}\lambda(4)^{Dd}
  \doteq\langle 1_a2_b3_c4_d \rangle,
\end{eqnarray}
\begin{eqnarray}
  \epsilon^{ABCD}
  \tilde{\lambda}(1)_{A}^{\dot{a}}
  \tilde{\lambda}(2)_{B}^{\dot{b}}
  \tilde{\lambda}(3)_{C}^{\dot{c}}
  \tilde{\lambda}(4)_{D}^{\dot{d}}
  \doteq [ 1_{\dot{a}}2_{\dot{b}}3_{\dot{c}}4_{\dot{d}} ].
\end{eqnarray}
One can also contract the momenta and spinors in the Lorentz invariant combinations:
\begin{eqnarray}
  \lambda(i)^{A_1a}p_{1,A_1A_2}p_{2}^{A_2A_3}
  \ldots p_{2n+1,A_{2n}A_{2n+1}}\lambda(j)^{A_{2n+1}b}\doteq
  \langle i^a |p_1p_2 \ldots p_{2n+1}|j^b\rangle
\end{eqnarray}
\begin{eqnarray}
  \lambda(i)^{A_1a}p_{1,A_1A_2}p_{2}^{A_2A_3}
  \ldots p^{A_{2n-1}A_{2n}}_{2n}\tilde{\lambda}(j)_{A_{2n}}^{\dot{b}}\doteq
   \langle i^a |p_1p_2 \ldots p_{2n}|j^{\dot{b}}]
\end{eqnarray}
Using the spinor products one can write, for example, an explicit expression for the $D=6$
gluon  polarization vector with momentum $p^{\mu}$ as:
\begin{eqnarray}
  \epsilon^{AB}_{a\dot{b}}(p)=
  \left(\lambda(p)^A_a\lambda(q)^B_b-
  \lambda(p)^B_a\lambda(q)^A_b\right)
  \langle q_b|p^{\dot{b}}]^{-1},
\end{eqnarray}
or equivalently
\begin{eqnarray}
  \epsilon_{a\dot{b}AB}(p)=
  \langle p^a|q_{\dot{c}}]^{-1}
  \left(\tilde{\lambda}(q)_{A\dot{c}}
  \tilde{\lambda}(p)_{B\dot{a}}-
  \tilde{\lambda}(q)_{B}^{\dot{c}}
  \tilde{\lambda}(p)_{A}^{\dot{a}}\right),
\end{eqnarray}
where it is implemented that the inverse matrices $\langle q_b|p^{\dot{b}}]^{-1}$
and $\langle p^a|q_{\dot{c}}]^{-1}$ are nondegenerate, which can be achieved
by an appropriate choice of reference momenta $q$ and associated spinors
$\lambda(q)^A_a$ and
$\tilde{\lambda}(q)_{A}^{\dot{a}}$.

Consider now the essential parts of the $D=6$ $\mathcal{N}=(1,1)$
on-shell momentum superspace construction. Some aspects of $\mathcal{N}=(1,1)$ $D=4,6$ gauge theories in the standard coordinate off-shell superspace have been  investigated in \cite{Ivanov_11_SUSY}. The on-shell
$\mathcal{N}=(1,1)$  superspace for $D=6$ SYM was first formulated in \cite{Sigel_D=6Formalism}. 
It can be parameterized by the following
set of coordinates
\begin{eqnarray}\label{Full_(1,1)_superspace}
  \mbox{$\mathcal{N}=(1,1)$ D=6 on-shell superspace}=\{\lambda^A_a,\tilde{\lambda}_{A}^{\dot{a}},\eta_a^I,\overline{\eta}_{I'\dot{a}}\},
\end{eqnarray}
where $\eta_a^I$ and $\overline{\eta}_{\dot{a}}^{I'}$ are the Grassmannian coordinates,
$I=1,2$ and $I'=1',2'$ are the $SU(2)_R\times SU(2)_R$ R-symmetry indices. Note that
this superspace is not chiral. We have two types of supercharges $q^{A I}$ and
$\overline{q}_{A I'}$ with the commutation relations
\begin{eqnarray}\label{commutators_for_superchrges_full_(1,1)}
  \{ q^{A I}, q^{B J}\}&=&p^{AB}\epsilon^{IJ},\nonumber\\
  \{ \overline{q}_{A I'}, \overline{q}_{B J'}\}&=&p_{AB}\epsilon_{I'J'},\nonumber\\
  \{ q^{A I},  \overline{q}_{B J'}\}&=&0.
\end{eqnarray}

The $\mathcal{N}=(1,1)$ $D=6$ SYM on-shell supermultiplet consists
of the gluon $A_{a\dot{a}}$, two fermions
$\Psi^a_I$,$\overline{\Psi}^{I'\dot{a}}$ and two complex scalars
$\phi^{I'}_I$ (antisymmetric with respect to $I,I'$). It is CPT
self-conjugated. However, to combine all the on-shell states in one
superstate $|\Omega\rangle$ by analogy with the $\mathcal{N}=4$
$D=4$ SYM one has to perform a truncation of the full
$\mathcal{N}=(1,1)$ on-shell superspace \cite{Sigel_D=6Formalism} in
contrast to the former case. Indeed, if one expands any function $X$
(or $|\Omega\rangle$ superstate) defined on the full on-shell
superspace in Grassmannian variables, one encounters  terms like
$\sim\eta_a^I\overline{\eta}_{I'\dot{a}}A_{I}^{I'a\dot{a}}$. Since
there are no such bosonic states $A_{I}^{I'a\dot{a}}$ in the
$\mathcal{N}=(1,1)$ SYM supermultiplet,  one needs to eliminate these
terms by imposing constraints on $X$, i.e.,  to truncate the full
on-shell superspace. If one wishes to use the little group indices
to label the on-shell states, the truncation has to be done with
respect to R symmetry indices. This can be done by consistently
using the harmonic superspace techniques \cite{Sigel_D=6Formalism}.
Defining the harmonic variables $u_{I}^{\mp}$ and $\overline{u}^{\pm
I'}$ which parametrize the double coset space
\begin{eqnarray}
\frac{SU(2)_R}{U(1)}\times\frac{SU(2)_R}{U(1)}
\end{eqnarray}
we express  the projected supercharges, the Grassmannian coordinates
\begin{eqnarray}
  q^{\mp A}&=&u^{\mp}_Iq^{A I},~
  \overline{q}^{\pm}_{A}=u^{\pm I'}\overline{q}_{A I'},
  \nonumber\\
  \eta^{\mp}_a&=&u^{\mp}_I\eta_a^I,~
  \overline{\eta}^{\pm}_{\dot{a}}=u^{\pm I'}\overline{\eta}_{I'\dot{a}},
\end{eqnarray}
and creation/annihilation operators of the on-shell states
\begin{eqnarray}\label{(1,1)_onshel_states}
  &&\phi^{--},~\phi^{-+},~\phi^{+-},~\phi^{++},\nonumber\\
  &&\Psi^{-a},~\Psi^{+a},~\overline{\Psi}^{-\dot{a}}~\overline{\Psi}^{+\dot{a}},\nonumber\\
  &&A^{a\dot{a}}.
\end{eqnarray}
in terms of the new harmonic variables.

Now one has to consider only the objects $X$ that depend
on half of the Grassmannian coordinates $\eta^{-}_a,\overline{\eta}_{\dot{a}}^{+}$, i.e.,
to impose the Grassmannian analyticity constraints on $X$:
\begin{eqnarray}
  D_{A}^+X=\overline{D}^{-A}X=0,
\end{eqnarray}
where $X$ is some function of the full on-shell superspace, and the projectors $D_{A}^{\pm}$ and
$\overline{D}^{\pm A}$ are  the
super covariant derivatives with respect to the supercharges
(\ref{commutators_for_superchrges_full_(1,1)}).
This can be done in a self consistent way if the projectors obey the algebra
\cite{Sigel_D=6Formalism}:
\begin{eqnarray}
  \{D_{A}^+,D_{B}^+\}=\{ \overline{D}^{-A},\overline{D}^{-B}\}=
\{D_{A}^+,\overline{D}^{-B}\}=0,
\end{eqnarray}
which is indeed the case due to
eq.(\ref{commutators_for_superchrges_full_(1,1)}). Therefore, in
what follows we will consider only objects that depend on the set of
variables which parametrize the subspace ("analytic superspace") of
the full $\mathcal{N}=(1,1)$ on-shell superspace
\begin{eqnarray}\label{Truncated_onshell_superspace}
  \{\lambda^A_a,\tilde{\lambda}_{A}^{\dot{a}},
\eta^{-}_a,\overline{\eta}_{\dot{a}}^{+} \}.
\end{eqnarray}
The projected supercharges acting on the analytic superspace
for the one particle case can be explicitly written as:
\begin{eqnarray}\label{projected_supercharges}
  q^{-A}=\lambda^A_a\eta^{-a},~
  \overline{q}_A^{+}=\tilde{\lambda}_A^{\dot{a}}\overline{\eta}_{\dot{a}}^{+}.
\end{eqnarray}
Now one can combine all the on-shell state creation/annihilation operators
(\ref{(1,1)_onshel_states}) into
one superstate $|\Omega_i\rangle=\Omega_i|0\rangle$ (here $i$ labels the momenta
carried by the state):
\begin{eqnarray}
  |\Omega_i\rangle&=&\{ \phi^{-+}_i+\phi^{++}_i(\eta^-\eta^-)_i+
  \phi^{--}_i(\overline{\eta}^+\overline{\eta}^+)_i
  +\phi^{+-}_i(\eta^-\eta^-)_i(\overline{\eta}^+\overline{\eta}^+)_i
  \nonumber\\
  &+&(\Psi^+\eta^-)_i+(\overline{\Psi}^-\overline{\eta}^+)_i+
  (\Psi^-\eta^-)_i(\overline{\eta}^+\overline{\eta}^+)_i+
  (\overline{\Psi}^+\overline{\eta}^+)_i(\eta^-\eta^-)_i\nonumber\\
  &+&(A\eta^-\overline{\eta}^+)_i\}|0\rangle,
\end{eqnarray}
where $(XY)_i\doteq X^{a/\dot{a}}_iY_{i~a/\dot{a}}$. Hereafter we
will drop the $\pm$ labels for simplicity.

Consider now the colour ordered n-particle superamplitude in the planar limit. The planar limit for
the $SU(N_c)$ gauge group  is understood as usual as the limit when $N_c\to\infty, g^2_{YM}\to 0$ and
$\lambda=g^2_{YM}N_c$ is fixed. We want to note that strictly speaking $g_{YM}$
is dimensional, so the real PT parameter would be $g_{YM}E$, where $E$ is some
energy scale. We will see later the explicit form of $E$.
We also use the "all ingoing notation" as usual. Then one has
\footnote{We do not write the S-matrix operator in the definition of $A_n$ explicitly. }
\begin{eqnarray}
 A_n(\{\lambda^A_a,\tilde{\lambda}_{A}^{\dot{a}},
\eta_a,\overline{\eta}_{\dot{a}} \})=\langle
0|\prod_{i=1}^n\Omega_i|0\rangle.
\end{eqnarray}
The superamplitude should be translationally invariant, i.e.,
it should be invariant under the action of supercharges (\ref{projected_supercharges}) and the total
momenta, i.e., translationally invariant in "analytic superspace". This means that
\begin{eqnarray}
 q^AA_n=\overline{q}_AA_n=p^{AB}A_n=0,
\end{eqnarray}
where for the n-particle case one has:
\begin{eqnarray}\label{projected_supercharges_n_particle_state}
  q^{A}=\sum_i^n\lambda^A_a(i)\eta^{a}_i,~
  \overline{q}_A=\sum_i^n\tilde{\lambda}_A^{\dot{a}}(i)\overline{\eta}_{\dot{a},i},
  ~p^{AB}=\sum_i^n\lambda^{Aa}(i)\lambda^B_{a}(i).
\end{eqnarray}
From the later we conclude that the superamplitude should have the form:
\begin{eqnarray}
 A_n(\{\lambda^A_a,\tilde{\lambda}_{A}^{\dot{a}},
\eta_a,\overline{\eta}_{\dot{a}} \})=
\delta^6(p^{AB})\delta^4(q^A)\delta^4(\overline{q}_A)\mathcal{P}_n(\{\lambda^A_a,\tilde{\lambda}_{A}^{\dot{a}},
\eta_a,\overline{\eta}_{\dot{a}} \}),
\end{eqnarray}
where $\mathcal{P}_n$ is a polynomial with respect to
$\eta$ and $\overline{\eta}$ of degree of $2n-8$. We will
drop the momentum conservation delta function $\delta^6(p^{AB})$
from now on. Note that since there is no helicity as a conserved quantum number,
there are no closed subsets of MHV, NMHV, etc. amplitudes,
and this is all we can say about the general structure
of the superamplitude from the supersymmetry alone.
The Grassmannian  delta functions $\delta^4(q^A)$ and $\delta^4(\overline{q}_A)$ are defined
in this case as:
\begin{eqnarray}
  \delta^4(q^A)&=&\frac{1}{4!}\epsilon_{ABCD}
  \delta(\sum_i^n q_i^A)\delta(\sum_k^n q_k^A)
  \delta(\sum_p^n q_p^A)\delta(\sum_l^n q_l^A),\nonumber\\
  \delta^4(\overline{q}_A)&=&\frac{1}{4!}\epsilon^{ABCD}
  \delta(\sum_i^n\overline{q}_{A,i})\delta(\sum_k^n\overline{q}_{A,k})
  \delta(\sum_p^n\overline{q}_{A,p})\delta(\sum_l^n\overline{q}_{A,l}).
\end{eqnarray}
The integral over $\delta(X)\delta(\overline{X})$ is performed according to the rule:
\begin{eqnarray}
  \int d \eta^a_i~\int d\overline{\eta}^{\dot{b}}_j~
  \delta(q^A)\delta(\overline{q}_{B})=\lambda(i)^{Aa}\tilde{\lambda}(j)_B^{\dot{b}},
\end{eqnarray}
so for the integral over the full superspace (\ref{Truncated_onshell_superspace}) is
\begin{eqnarray}
  \int d^2 \eta^a_{l_1}~\int d^2\overline{\eta}^{\dot{b}}_{l_2}
  \doteq \int d^4\eta_{l_1l_2},
\end{eqnarray}
and we obtain\footnote{Note that we do not integrate over harmonics
$u^{\mp I}$ and $\overline{u}^{\pm I'}$, so in this formulation R
symmetry is not explicit \cite{Sigel_D=6Formalism}. We choose different from \cite{Brandhuber_1loopD=6}
normalization of the supersum. This can be done by rescaling of
$g_{YM}^2$ by $1/32$.}:
\begin{eqnarray}\label{Two_particle_Grassmann_int}
  \int d^4\eta_{l_1l_2}d^4\eta_{l_2l_1}~
  \delta^4(q^A)\delta^4(\overline{q}_{B})=2(l_1,l_2)^2.
\end{eqnarray}
This is an important relation because it
allows one to compute sums over the states ("supersums") which appear in
two particle cuts within the unitarity based computations. As we
will see in the next section for the 4-point amplitude the two
particle iterated cuts will be sufficient to construct the integrand
up to two loops, just as in the $D=4$ $\mathcal{N}=4$ SYM case.

\section{The 4-point amplitude in D=6}\label{4point ampl_3}
Consider the simplest amplitudes with 4 legs.
For $n=4$ the degree of Grassmannian polynomial $\mathcal{P}_4$ is
$2n-8=0$, so $\mathcal{P}_4$ is a function of bosonic variables
$\{\lambda^A_a,\tilde{\lambda}_{A}^{\dot{a}}\}$ only
\begin{eqnarray}\label{4-point_ampl_general_form}
  A_4(\{\lambda^A_a,\tilde{\lambda}_{A}^{\dot{a}},
\eta_a,\overline{\eta}_{\dot{a}} \})=
  \delta^4(q^A)\delta^4(\overline{q}_A)
  \mathcal{P}_4(\{\lambda^A_a,\tilde{\lambda}_{A}^{\dot{a}}\}).
\end{eqnarray}
At the tree level $\mathcal{P}_4$ can be found from the explicit answer in components
for the 4 gluon amplitude \cite{DonaldOConnel_AmplInD=6,Sigel_D=6Formalism}
obtained by using the six dimensional version of the BCFW recurrence relation:
\begin{eqnarray}
  \mathcal{A}_4^{(0)}(1_{a\dot{a}}2_{b\dot{b}}3_{c\dot{c}}4_{d\dot{d}})=
  -ig_{YM}^2
  \frac{\langle
  1_a2_b3_c4_d\rangle[1_{\dot{a}}2_{\dot{b}}3_{\dot{c}}4_{\dot{d}}]}{st}.
\end{eqnarray}
Comparing this expression with (\ref{4-point_ampl_general_form}) and
expanding (\ref{4-point_ampl_general_form}) in powers of $\eta$, $\overline{\eta}$
and then extracting the coefficient of
$(\eta\overline{\eta})_1(\eta\overline{\eta})_2(\eta\overline{\eta})_3(\eta\overline{\eta})_4$
one concludes that:
$$\mathcal{P}_4^{(0)}=-ig_{YM}^2/st,$$
where $s$ and $t$ are the standard Mandelstam
variables\footnote{We use the
compact notation for scalar products of
massless momenta $(p_i+p_j)^2=(i+j)^2=2(ij)$.} $s=(1+2)^2$, $t=(2+3)^2$. So, at the tree level the 4-point
superamplitude can be written in a very compact form:
\begin{eqnarray}\label{4_point_tree_superamplitude}
  A_4^{(0)}=-ig_{YM}^2\frac{\delta^4(q^A)\delta^4(\overline{q}_A)}{st}.
\end{eqnarray}
Note that already at the tree level the 5-point amplitude is not so simple
\cite{DonaldOConnel_AmplInD=6,Sigel_D=6Formalism}.

The $D=6$ SYM theory has the dimensional coupling constant and it is not conformal
already at the classical level. On other hand, the tree level amplitudes in this
theory possess the  dual conformal covariance \cite{D6_DualConformal_Invariance} ,
i.e., they are covariant under inversions $I$
and special conformal transformations $K^{\mu}$.
The explicit form of $I$, $K^{\mu}$ and the way how the dual coordinates are introduced
can be found in \cite{D6_DualConformal_Invariance}.
Under these transformations of momenta  the amplitudes
transform as
\begin{eqnarray}
  I[\mathcal{P}_n^{(0)}]&=&\prod_{i=1}^nx_i^2\mathcal{P}_n^{(0)},\nonumber\\
  K^{\mu}[\mathcal{P}_n^{(0)}]&=&\sum_{i=1}^n2x^{\mu}_i\mathcal{P}_n^{(0)},
\end{eqnarray}
where $x_i^{\mu}$ are the dual coordinates for momenta $p^{\mu}_i$. The combination of $\delta$-functions
$\delta^6(p^{AB})\delta^4(q^A)\delta^4(\overline{q}_A)$ transforms covariantly
with the factor $(x_1^2)^2$; we will not write them hereafter.
 At the loop level the L-loop integrand $Int\mathcal{P}^{(L)}_n$ transforms under
inversions $I$ as:
\begin{eqnarray}
  I[Int\mathcal{P}^{(L)}_n]=\prod_{i=1}^n x^2_i\prod_{k=1}^L (x_{l_k}^2)^4
  Int\mathcal{P}^{(L)}_n,
\end{eqnarray}
where $l_k$ is the loop momenta. So the loop amplitude is dually
conformal covariant if the loop momentum integration $d^{D=6}l_i$ is
restricted to the $d^{D=4}l_i$ subspace only.

It seams that despite the presence of the dual  conformal
invariance/covariance in the amplitudes of higher dimensional gauge
theories its true power at the loop level manifests itself only in four
dimensions. Intuitively one may think that it should work in the
opposite way.

In the case of the $D=4$ $\mathcal{N}=4$ SYM, the dual conformal
invariance plays an extremely important role. For example, the whole
existence of the exact BDS formulas for the 4- and 5-point MHV
amplitudes can be understood as a consequence of the dual conformal
invariance. In the case of $D=6$ $\mathcal{N}=(1,1)$ SYM we do not
expect similar all loop restrictions but in both theories the dual
conformal invariance still restricts the form of the integrands.

Let us now make some short comments about the UV/IR structure of the
$D=6$ $\mathcal{N}=(1,1)$ SYM amplitudes. In \cite{2loopN=4 and
finitenes bound 07 Bern&Co,N=4finitenessBound}, the UV finiteness
bound
\begin{eqnarray}
  D < 4+\frac{6}{L},~L\geq2.
\end{eqnarray}
was suggested for the gauge theory amplitudes with maximal
supersymmetry. The one loop level is exceptional and the first UV
divergences may appear at D=8 and not at D=10
\cite{1loopFinitnesBoud}. We see that for $\mathcal{N}=(1,1)$ SYM at
three loops $D=6$ is a critical dimension in a sense that the first
UV divergence may appear. This is consistent with the old estimates
based on off-shell superspace considerations that suggest that the
first UV divergence may appear after three loops, i.e.
$\mathcal{N}=(1,1)$ SYM is one and two loop finite \cite{(11)
finiteness at 2 loops}. As we will see in a moment this is just the
case. In recent years, interest in the  UV properties of the
S-matrix of formally nonrenormalizable gauge theories with extended
supersymmetry ($D=4$ $\mathcal{N}=8$ SUGRA is a particular example)
was reborn. One may hope that there are theories with the UV finite
S-matrix which are formally nonrenormalizable \cite{N=8SUGRA
finiteness}. The results obtained so far are in some sense
controversial \cite{SUGRA fin Vs Div,D=5SYM_Diverges_ZBernDixon} but
one may  still hope for the UV finiteness of $D=4$ $\mathcal{N}=8$
SUGRA. Regardless of these results we treat the $D=6$
$\mathcal{N}=(1,1)$ SYM amplitudes in the following way. We compute
the 4-point amplitude at the two loop approximation and study the
high energy (Regge) asymptotics at all loops. In our
considerations we do not encounter any UV divergences. If $D=6$
$\mathcal{N}=(1,1)$ SYM at higher orders is not UV finite, still one
can consider the high energy (Regge) asymptotic behaviour of the
n-point amplitudes as computation of some particular limit of the
corresponding string/M theory S-matrix. It is interesting to note
that there are no IR divergences in $D=6$ $\mathcal{N}=(1,1)$ SYM,
so we obtain completely \emph{finite} answers for the amplitudes,
contrary to the $D=4$ $\mathcal{N}=4$ case.

\begin{figure}[ht]
 \begin{center}
  \epsfxsize=6cm
 \epsffile{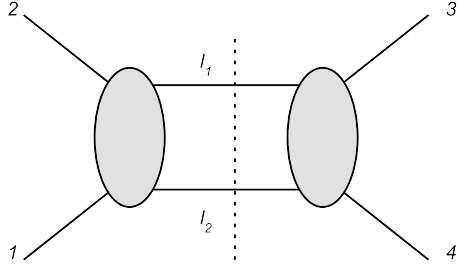}
 \end{center}\vspace{-0.2cm}
 \caption{Two particle s-channel cut for the one loop amplitude.}\label{1loop2particle}
 \end{figure}

Consider now the structure of the four point amplitude at one and two loops. In the unitarity
based approach this computation is essentially trivial. Since we expect
no UV/IR divergences up to two loops,  the amplitudes at this
order of PT can be obtained by the unitary cut method without any regularization. The
easiest way to obtain the answers in terms of scalar integrals is to use the super amplitude
(\ref{4_point_tree_superamplitude}) and impose the two particle cuts. We use the notation
\begin{eqnarray}
  q^A_L&=&q^A_1+q^A_2,\nonumber\\
  q^A_R&=&q^A_3+q^A_4,\nonumber\\
  q^A_{l_1l_2}&=&q^A_{l_1}+q^A_{l_2}.
\end{eqnarray}
and assume the momentum conservation conditions associated
with the amplitudes on both sides of the cut: $1+2+l_1+l_2=0$ and
$-l_1-l_2+3+4=0$ (see Fig.\ref{1loop2particle}).
 The integrand for the  s-channel two particle cut of the one loop amplitude takes the form
\cite{Sigel_D=6Formalism,Brandhuber_1loopD=6}
\begin{eqnarray}
  IntA_4^{(1)}&=&\int d^4\eta_{l_1l_2}d^4\eta_{l_2l_1}~
  A^{(0)}_4(1,2,l_1,l_2)\times A^{(0)}_4(-l_1,-l_2,3,4)
 \end{eqnarray}
 Using (\ref{4_point_tree_superamplitude}) and (\ref{Two_particle_Grassmann_int})
and momentum conservation conditions one gets:
(the common factor $g_{YM}^4 N_c$ is omitted)
\begin{eqnarray}
 IntA_4^{(1)}
&=&-\int d^4\eta_{l_1l_2}d^4\eta_{l_2l_1}
\frac{\delta^4(q^A_R+q^A_{l_1l_2})\delta^4(q^A_L-q^A_{l_1l_2})
\delta^4(\overline{q}_{A,R}+\overline{q}_{A,l_1l_2})
\delta^4(\overline{q}_{A,L}-\overline{q}_{A,l_1l_2})}{s^2(2+l_1)^2(4+l_2)^2}
\nonumber\\
&=&-\delta^4(q^A_R+q^A_L)\delta^4(\overline{q}_{A,R}+\overline{q}_{A,L})
\frac{2(l_1l_2)^2}{s^2(2+l_1)^2(4+l_2)^2}=A^{(0)}_4\frac{st}{2}
\frac{-i}{(2+l_1)^2(4+l_2)^2},\nonumber\\
\end{eqnarray}
which is consistent with the  following ansatz for
part of the amplitude associated with the s-channel cut
\begin{eqnarray}
  -A^{(0)}_4~\frac{st}{2}~B(s,t),
\end{eqnarray}
where $B(s,t)$ is the $D=6$ scalar box function.
The t-channel cut gives the same result, so we conclude that the full one loop level
amplitude has the form:
\begin{eqnarray}\label{1-loop 4point ampl scalar int}
    A_4^{(1)}=-A^{(0)}_4~\frac{g_{YM}^2 N_c}{2}~st~B(s,t).
\end{eqnarray}

\begin{figure}[ht]
 \begin{center}
  \epsfxsize=6cm
 \epsffile{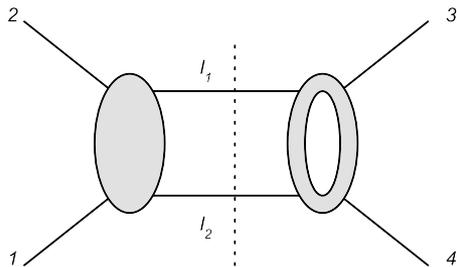}
 \end{center}\vspace{-0.2cm}
 \caption{Two particle s-channel cut for the two loop amplitude.}\label{2loop2particle}
 \end{figure}

\begin{figure}[t]
 \begin{center}
  \epsfxsize=10cm
 \epsffile{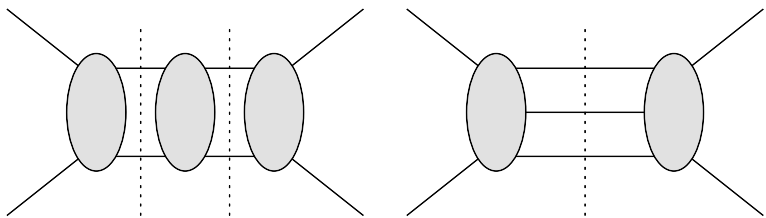}
 \end{center}\vspace{-0.2cm}
 \caption{Two loop iterated two particle and three particle cuts.}\label{2loopIter3part}
 \end{figure}

Consider now the two loops. Applying the two particle cut
for the s-channel  in a similar way  as at the one loop level one gets
(see Fig.\ref{2loop2particle})
\begin{eqnarray}
  IntA_4^{(2)}&=&\int d^4\eta_{l_1l_2}d^4\eta_{l_2l_1}~
  A^{(0)}_4(1,2,l_1,l_2)\times A^{(1)}_4(-l_1,-l_2,3,4)
  \nonumber\\
&=&\int d^4\eta_{l_1l_2}d^4\eta_{l_2l_1}
\frac{\delta^4(q^A_R+q^A_{l_1l_2})\delta^4(q^A_L-q^A_{l_1l_2})}{2s}
\nonumber\\
&\times&\frac{\delta^4(\overline{q}_{A,R}+\overline{q}_{A,l_1l_2})
\delta^4(\overline{q}_{A,L}-\overline{q}_{A,l_1l_2})}{(2+l_1)^2}B(s,(4+l_2)^2)
\nonumber\\
&=&\delta^4(q^A_R+q^A_L)\delta^4(\overline{q}_{A,R}+\overline{q}_{A,L})
\frac{(l_1l_2)^2}{s(2+l_1)^2}B(s,(4+l_2)^2)
\nonumber\\
&=&A^{(0)}_4~\frac{s^2t}{4}~\frac{i}{(2+l_1)^2}B(s,(4+l_2)^2),
\end{eqnarray}
which is consistent with the  following ansatz for
part of the amplitude associated with the s-channel cut
\begin{eqnarray}
  A_4^{(2)}|_s=A^{(0)}_4~\frac{s^2t}{4}~DB(s,t),
\end{eqnarray}
where $DB(t,s)$ is the $D=6$ scalar double box function. The t-channel two
particle cut gives a similar result:
\begin{eqnarray}
  A_4^{(2)}|_t=A^{(0)}_4~\frac{st^2}{4}~DB(t,s),
\end{eqnarray}
so combining the two contributions together we obtain:
\begin{eqnarray}\label{2-loop 4point ampl scalar int}
  A_4^{(2)}=A^{(0)}_4~\frac{(g_{YM}^2 N_c)^2}{4}\left(s^2t~DB(s,t)+st^2~DB(t,s)\right).
\end{eqnarray}
The three particle cuts \cite{ZBern_GenUnit_D=6Helicity} and the
double two particle cuts which are schematically presented in
Fig.(\ref{2loopIter3part}) do not give any new information. One can
see that in fact the answers for the amplitude in terms of the
scalar integrals are identical to those of the  $D=4$
$\mathcal{N}=4$ SYM \cite{ZBern_GenUnit_D=6Helicity} with exchange
of the loop momentum integrations from $D=4$ to $D=6$. This may be
understood as a consequence of a dual conformal covariance of the
integrands. Also note that at the L loop level we always encounter
the L-loop scalar L-rang ladder integrals.

\section{The six dimensional boxes and the high energy limit}\label{BoxInt by MB_4}
The $D=6$ scalar boxes $B(s,t)$ and $DB(s,t)$ are completely finite functions of the
Mandelstam variables and can be computed in terms of logarithms, Polylogarithms
and harmonic sums by means of either the Feynman parametrization or the MB representation
technique (see appendix).

The evaluation of the single  box $B(s,t)$ is streitforward and gives
\begin{eqnarray}\label{b2}
 B(s,t)=\frac{\pi^{3}}{(2\pi)^6}\ \frac{b_2(x)}{s+t},\ \ ~b_2(x)=\frac{L^2(x)+\pi^2}{2},\ \ ~x=\frac{t}{s}.
\end{eqnarray}
where $L(x)\doteq\log(x)$.

The double box $DB(s,t)$ can be evaluated by means of the MB representation. We reproduce this derivation in Appendix A.
One has \cite{Tausk}:
\begin{eqnarray}\label{DB}
  DB(s,t)=\left(\frac{\pi^3}{(2\pi)^6}\right)^2\left(\frac{b_4(x)}{t}+\frac{b_3(x)}{s+t}\right),
\end{eqnarray}
where
\begin{eqnarray}\label{b4}
  b_4(x)&=&\left(2\zeta_3-2Li_3(-x)-\frac{\pi^2}{3}L(x)\right)L(1+x)
  +\left(\frac{1}{2}L(x)+\frac{\pi^2}{2}\right)L^2(1+x)\nonumber\\
  &+&\left(2L(x)L(1+x)-\frac{\pi^2}{3}\right)Li_2(-x)+2L(x)S_{1,2}(-x)-2S_{2,2}(-x),
\end{eqnarray}
\begin{eqnarray}\label{b3}
  b_3(x)=-2\zeta_3+\frac{\pi^2}{3}L(x)-\left(L(x)+\pi^2\right)L(1+x)-2L(x)Li_2(-x)+2Li_3(-x)
\end{eqnarray}
and $S_{1,2}$ and $S_{2,2}$ are the harmonic polynomials. Note that both
expressions for $B(s,t)$ and $DB(s,t)$ are real when $s>0,t>0$ (euclidian region), while
in the physical region $t<0$ the imaginary part appears.

The expression for the  double box (\ref{DB}-\ref{b3})  contains all
typical transcendental structures and does not reduce to logarithms
contrary to the 4-point function in the $D=4$ $\mathcal{N}=4$ case.
This does not happen for the full answer (\ref{2-loop 4point ampl
scalar int}) as well, the Polylog functions remain. Note that even
for the full amplitude in contrast to the $D=4$ $\mathcal{N}=4$ SYM
case the maximal transcendentality principle no longer holds. While
both $b_2(x)$ and $b_4(x)$ are uniform and obey the maximal
transcendentality criteria\footnote{If we attach to each logarithm
and $\pi$ the level of transcendentality equal to $1$ and to
Polylogarithms $Li_n(x)$ and $\zeta_n$ the level of
transcendentality equal to $n$, then  at the given order of
perturbation theory the coefficient for the $n$-th pole
$1/\epsilon^n$ has the overall transcendentality  equal to $2l-n$,
where $l$ is the number of loops; $n=0$ corresponds to a finite
part. For a product of several factors it is given by the  sum of
transcendentalities  of each factor.}, $b_3(x)$ is also uniform but
has a lower transcedentality level.

In the case of $D=4$ $\mathcal{N}=4$ SYM there is a conjecture  that
the maximal transcendentality principle can be explained by using
recurrence relations for the integrands written in terms of momentum
twistor variables \cite{Arcani-Hamed Grassmannians}. One may
speculate that some generalization of the maximal transcendentality
principle still holds for the $D=6$ $\mathcal{N}=(1,1)$ SYM.  Indeed,
in the $D=6$ case, there exists a supertwistor formalism based on
$OSp^{*}(8|2)$ superconformal group \cite{Sigel_D=6Formalism}. Also,
the $D=6$ $\mathcal{N}=(1,1)$ SYM amplitudes at the tree level can
be understood as the Higgs regulated $D=4$ $\mathcal{N}=4$ SYM ones
and the construction of the integrand is based only on properties of
the tree level amplitudes.

Let us now consider the high energy behaviour of our four point amplitude. In this
regime it is usually possible to obtain simple  expressions in each order
of PT so that the all loop summation in terms of known functions becomes
possible.  One can also think of the high energy behaviour of the field theory amplitudes
as a special limit of the corresponding string theory S-matrix.

Consider the so-called  Regge  limit for the four point amplitude.
In the Regge limit when $s\to+\infty$  and $t<0$ is fixed the main contribution
comes from the vertical ladder diagrams. At the one loop order  one has:
\begin{eqnarray}\label{l2}
B(s,t)|_{s\to\infty}\simeq \frac{1}{2} \frac{ L^2(x) }{s}+... .
\end{eqnarray}
At two loops the main contribution comes from the
vertical double box $DB(t,s)$ which is equal
to
\begin{eqnarray}\label{l4}
DB(t,s)|_{s\to\infty}\simeq \frac{1}{12} \frac{ L^4(x) }{s}+... .
\end{eqnarray}
We neglect here all the terms $\sim L^{k}(x)$, with $k<2n$ and the constants.

Substituting eqs.(\ref{l2}) and (\ref{l4}) into eq.(\ref{2-loop 4point ampl scalar int}) one gets
\begin{equation}
  A_4\simeq A^{(0)}_4\left[1+\frac{g_{YM}^2 N_c|t|}{128\pi^3}\frac{L^2(x)}{2}
  +\left(\frac{g_{YM}^2 N_c|t|}{128\pi^3}\right)^2\frac{L^4(x)}{12}+... \right].
\end{equation}
Note that the dimensional coupling  $g_{YM}^2$ is always multiplied by
$t$ forming the dimensionless expansion parameter $g_{YM}^2N_c|t|$.

In higher orders of PT the main contribution in this limit also comes from
the vertical multiple boxes, the so-called ladder diagrams.
Their asymptotics are well known in $D=4$ and can be similarly evaluated
in $D=6$. We consider this derivation in Appendix B.
The result for the Regge limit of the vertical n-loop ladder diagram is
UV and IR finite and takes the form
\begin{equation}\label{L-rung box assympt}
\frac 1s \frac{L^{2n}(x)}{n!(n+1)!}.
\end{equation}
Combined with the combinatorial factor $s t^n/2^n$  this leads to the series
\begin{equation}
A_4\simeq A^{(0)}_4 \sum_{n=0}^{\infty} \frac{\lambda^n L^{2n}(x)}{n!(n+1)!},
\ \ \ \lambda\equiv\frac{g_{YM}^2 N_c|t|}{128\pi^3}.
\end{equation}
This series can be summed and represents the Bessel function of the imaginary argument
\begin{equation}\label{AmplitudeBessel}
A_4\simeq A^{(0)}_4 \frac{I_1(2 y)}{y}, \ \ \ y\equiv\sqrt{\lambda}
L(x).
\end{equation}
In the Regge limit when $y\to\infty$ $I_1(2 y)\to \exp(2y)/(2\sqrt{\pi y})$
and one gets the Regge type behaviour
\footnote{Note that the tree amplitude
$A_4^{(0)}\sim s/t$  just like in four dimensions}
\begin{eqnarray}
  \frac{A_4}{A_4^{(0)}}\sim\left(\frac{s}{t}\right)^{\alpha(t)-1}
\end{eqnarray}
with
\begin{eqnarray}
  \alpha(t)=1+2\sqrt{\lambda}=1+\sqrt{\frac{g_{YM}^2 N_c|t|}{32\pi^3}}~.
\end{eqnarray}
We want to stress once again that  all contributions
from the terms $\lambda^nL^k(x)$ with $k<2n$ are omitted. One can
see that as expected for the gauge theory $\alpha(0)=1$. Note that
because there are no UV/IR divergences in the Regge limit in the
$D=6$ $\mathcal{N}=(1,1)$ SYM our result for the amplitude is
completely independent of any kind of regulator. Notice also that
the limit $y\to\infty$ can be achieved in two regimes:
\begin{eqnarray}
 L(x)\gg1,~\lambda<1\ \ \ ~\mbox{or} \ \ \ \lambda > L(x) \gg 1.
\end{eqnarray}
The first one is the weak coupling regime while the second one
resembles the strong coupling limit.

It is interesting to compare this result to the Regge behaviour of
the $D=4$ gauge and gravity theories with maximal supersymmetry:
$\mathcal{N}=4$ SYM and $\mathcal{N}=8$ SUGRA. For the $D=4$
$\mathcal{N}=4$ SYM the exact expression for 4-point amplitude is given
by the BDS ansatz. In the Regge limit in dimensional regularization
the BDS ansatz reduces to \cite{Regge N=4 SYM,Regge N=8SUGRA N=4
SYM} (see also \cite{LipatovKotikovN=4} for the recent discussion):
\begin{eqnarray}
 \frac{A_4} {A_4^{(0)}}
 \sim\left(\frac{s}{t}\right)^{\alpha(t)-1},
\end{eqnarray}
with (we assume that $t \gg \mu^2$, where $\mu^2$ is the dimensional
parameter of the dimensional regularization parameter, and
$\lambda_{4}=g_{4}^2N_c$, where $g_{4}$ is the dimensionless  coupling
constant of the $D=4$ SYM theory)
\begin{eqnarray}
  \alpha(t)=1-\frac{f(\lambda_{4})}{4}L\left(\frac{t}{\mu^2}\right),
\end{eqnarray}
where $f(\lambda_4)$ is the cusp anomalous
dimension. In the weak/strong coupling regimes one has:
\begin{eqnarray}
  \alpha(t)&=&1-\frac{\lambda_{4}}{8\pi^2}
  L\left(\frac{t}{\mu^2}\right)+...,
  ~\lambda_{4} \ll1 \nonumber\\
    \alpha(t)&=&1-\left(\frac{\sqrt{\lambda_{4}}}{\pi}\right)
    L\left(\frac{t}{\mu^2}\right)+...,
    ~\lambda_{4}\gg 1.
\end{eqnarray}
It is remarkable that in $D=6$ $\mathcal{N}=(1,1)$ the dependence of
$\alpha(t)$ on the effective coupling $\lambda$ is similar to that
 in the $D=4$ $\mathcal{N}=4$
SYM in the strong coupling regime. Note also that the
result of summation of the leading logarithms (\ref{AmplitudeBessel}) is
similar to the exact result for the vacuum expectation of a circular
Wilson loop in the $D=4$ $\mathcal{N}=4$ SYM \cite{Circular Wilson Loop
Zarembo}.

For the $N=8$ SUGRA one has \cite{Regge N=8SUGRA N=4
SYM,LipatovKotikovN=8}:
\begin{eqnarray}
 \frac{A_4} {A_4^{(0)}}
 \sim\left(\frac{s}{t}\right)^{\alpha(t)-2},
\end{eqnarray}
with ($k$ is the dimensional $D=4$ gravitational coupling constant)
\begin{eqnarray}
  \alpha(t)=2-\frac{k^2t}{2}L\left(\frac{t}{\mu^2}\right)+...~.
\end{eqnarray}
The effective coupling constant  here is $k^2t$ like in the $D=6$
$\mathcal{N}=(1,1)$ SYM.

\section{Conclusion}\label{Conclusion_5}
In this article we discussed the structure of the four point
amplitude in the $D=6$ $\mathcal{N}=(1,1)$ SYM at one and two loop
orders  in the planar limit and studied the high energy asymptotics
in the Regge limit.

The reduction of the one and two loop amplitudes to the scalar
integrals is essentially trivial when the $D=6$ spinor helicity and
the on-shell momentum superspace formalisms are used. Up to two
loops all the scalar integrals can be written in terms of the box
and double box functions in $D=6$ which can be evaluated by the MB
representation method. These functions are IR and UV finite in
agreement with the UV finiteness bounds. The three loop computations
are also possible; however, they are more involved since
the Barnes lemmas are no more sufficient to compute the $D=6$ three loop
boxes. The answers for the one and two loop $D=6$ boxes can be
written in terms of logarithms, Polylogarithms and harmonic
polynomials (harmonic Polylogarithms) of transcendentality 2 and 4 or
3 at one and two loops, respectively.

We see that for the full amplitude the contributions with
transcendentality 3 do not cancel, so the maximal transcendentality
principle no longer holds. Still one may wonder whether some
generalization of the maximal transcendentality principle may be
formulated. Indeed, the integrands of $D=6$ $\mathcal{N}=(1,1)$ SYM
can be interpreted as integrands of $D=4$ $\mathcal{N}$=4 SYM on a Coulomb
branch (Higgs regulated). Also, for $D=6$ there exists a twistor
formalism based on the $OSp^{*}(8|2)$ superconformal group, which
may be useful in explicit computations as our experience with the
$D=4$ $\mathcal{N}$=4 SYM tells us.

The high energy limit ($s\gg 1$) of the four point amplitude is
determined by the contributions of the vertical $D=6$ L-rung boxes,
whose leading asymptotics  can be evaluated.  The all order
summation gives the Bessel function from which the Regge behaviour
of the amplitude with $\alpha(0)=1$ can be obtained as expected. It
is interesting to note a similar dependence of $\alpha(t)$ on the
effective coupling
 $\lambda$, as in the strong coupling limit of the $D=4$
$\mathcal{N}=4$ SYM.

In our analysis we completely ignored a possible nonperturbative
contribution from the classical field configurations. They might be
interesting  by themselves.  The instantons from $D=4$ when
uplifted to $D=6$ become
 instantonic strings, the
one dimensional objects with their own nontrivial dynamics.
It would be interesting to study how such contributions  might affect the scattering amplitudes.

Another interesting question is whether there is some form of
geometrical realization of symmetries of $D=6$ $\mathcal{N}=(1,1)$
SYM, i.e., some analog of the Wilson loop/amplitude duality for the
$D=6$ $\mathcal{N}=(1,1)$ SYM
\cite{SpinorHelisityForm_D=10Dimentions}. We hope that the results
obtained here might be useful in this quest.

\section{Appendix A}
Here we present the evaluation of the box and double box integrals.
The box integral is defined as
\begin{eqnarray}
Box(s,t)=\frac 1i\int \frac{d^6k}{(2\pi)^6}
  \frac{1}{k^2(k+p_1)^2(k+p_1+p_2)^2(k-p_4)^2}.
\end{eqnarray}
This integral can be easily evaluated by Feynman parametrization.
The result is given by (\ref{b2}).

The double box integral is defined as
\begin{eqnarray}
DBox(s,t)&=&\frac{1}{i^2}\int \frac{d^6k}{(2\pi)^6}\frac{d^6l}{(2\pi)^6}
  \frac{1}{k^2l^2(k+p_1)^2(k+p_1+p_2)^2}\nonumber\\
  &\times&\frac{1}{(l+p_1+p_2)^2(l+p_1+p_2+p_4)^2(k-l)^2}.
\end{eqnarray}
This integral is evaluated with the help of the Mellin-Barnes
representation method. We use the MB expression of the horizontal
double box integral from V. Smirnov's book \cite{Smirnov Book}
\begin{eqnarray}
DBox_6(s,t)\!\!\!\!&=&\!\!\!\!
\frac{-\pi^{6}}{s}\int_{-i\infty}^{i\infty}\!\!
\frac{dz_1...dz_4}{(2\pi)^{12}(2\pi i)^4} \ x^{z_1}
\frac{\Gamma(1+z_1)\Gamma(-z_1-z_2)\Gamma(-z_1-z_3)
\Gamma(-z_2-z_3-z_4)}{(z_2+z_4)(z_3+z_4)(2+z_1-z_4)(1+z_1+z_4)}\nonumber \\   && \nonumber\\
&&\hspace{0.1cm} \times \ \ \
\Gamma(-z_1)\Gamma(1+z_1+z_2+z_3+z_4)\Gamma(1+z_1-z_4)\Gamma(1+z_2)\Gamma(1+z_3)\Gamma(z_4),\nonumber\\
\label{MBDB}
\end{eqnarray}
where $x=t/s$. The Mellin-Barnes integrals can be evaluated by using the Barnes lemmas (see \cite{Smirnov Book}, Ch.D).
To control the correctness of the choice of the integration contour, we check each step numerically.
For this purpose one has to choose first the real parts of the integration variables $z_i$ in such a way that all the arguments of the $\Gamma$ functions in (\ref{MBDB}) are positive. One of the possible choices is
$z_1 = -1/4, z_2 = -9/32, z_3 = -27/64, z_4 = 9/16$. The result does not depend on a particular choice.

The integral over $z_2$ is straightforward with the help of the first lemma and the second lemmas
\begin{eqnarray}
&& \int_{-i\infty}^{i\infty} \frac{dz_2}{2\pi i} \ \frac{\Gamma(1+z_2)\Gamma(-z_1-z_2)
\Gamma(-z_2-z_3-z_4)\Gamma(1+z_1+z_2+z_3+z_4)}{(z_2+z_4)}\nonumber \\  &=&
\Gamma(1-z_1)\Gamma(z_3+z_4)\Gamma(z_1)\Gamma(1-z_3-z_4)\left(1-\frac{\Gamma(-z_1+z_4))\Gamma(-z_3)}{\Gamma(z_4)\Gamma(-z_1-z_3)}\right).
\end{eqnarray}

The integral over $z_3$ is already tricky due to the  degeneracy of the arguments of the $\Gamma$ functions, and one has to modify the contour to keep all the poles to the right. The result is
\begin{eqnarray}
&-& \int_{-i\infty}^{i\infty} \frac{dz_3}{2\pi i} \ \Gamma(1+z_3)\Gamma(-z_1-z_3)
\Gamma(-z_3-z_4)\Gamma(z_3+z_4)\left(1-\frac{\Gamma(-z_1+z_4)\Gamma(-z_3)}{\Gamma(z_4)\Gamma(-z_1-z_3)}\right)\nonumber
\\
&=&\Gamma(1-z_4)\Gamma(-z_1+z_4)\left[\psi(z_4)-\psi(-z_1+z_4)+\psi(1-z_1)-\psi(1)\right].
\end{eqnarray}

The integral over $z_4$ can be composed into 2 integrals which again can be evaluated with the help of derivative of the first lemma
$$\int_{-i\infty}^{i\infty}\! \frac{dz_4}{2\pi i} \
\frac{\Gamma(1-z_4)\Gamma(-z_1+z_4)\Gamma(1+z_1-z_4)\Gamma(z_4)}{(2+z_1-z_4)(1+z_1-z_4)}\left[\psi(z_4)\!-\!\psi(-z_1+z_4)\!+\!\psi(1-z_1)\!-\!\psi(1)\right]$$
$$=-\!\int_{-i\infty}^{i\infty}\!\! \frac{dz_4}{2\pi i} \
\Gamma(1-\!z_4)\Gamma(-\!1-\!z_1+\!z_4)\Gamma(1+z_1-z_4)\Gamma(z_4)\left[\psi(z_4)\!-\!\psi(-z_1+z_4)\!+\!\psi(1\!-\!z_1)\!-\!\psi(1)\right]$$
$$-\!\int_{-i\infty}^{i\infty}\! \frac{dz_4}{2\pi i} \
\Gamma(1-\!z_4)\Gamma(-\!2-z_1+\!z_4)\Gamma(2+z_1-z_4)\Gamma(z_4)\left[\psi(z_4)\!-\!\psi(-z_1+z_4)\!+\!\psi(1\!-\!z_1)\!-\!\psi(1)\right].$$
After a careful choice of the integration contours one has for each term separately:
\begin{eqnarray*}
I_{11}&=&-\Gamma(1+z_1)\Gamma(-z_1)\left[\psi^2(1+z_1)-\psi(1+z_1)\psi(1)+\psi'(1+z_1)-\psi'(1)\right],\\
I_{21}&=&-\Gamma(2+z_1)\Gamma(-1-z_1)\left[\psi^2(2+z_1)-\psi(2+z_1)\psi(1)+\psi'(2+z_1)-\psi'(1)\right.\\
&&\left.-\psi(1+z_1)\right],\\
I_{12}&=&\frac 12\Gamma(1+z_1)\Gamma(-z_1)\left[2\psi(1+z_1)\psi(1)-2\psi(-z_1)\psi(1)+\psi^2(-z_1)-\psi^2(1)\right.\\
&&\left. +\frac{\pi^2}{2}-\psi'(-z_1)\right],\\
I_{22}&=&\frac 12\Gamma(2+z_1)\Gamma(-1-z_1)\left[2\psi(2+z_1)\psi(1)-2\psi(-1-z_1)\psi(1)+\psi^2(-1-z_1)-\psi^2(1)\right. \\
&&\left.+\frac{\pi^2}{2} -\psi'(-1-z_1)+\frac{1}{1+z_1}+2\psi(1+z_1)-2\psi(-z_1)-2\psi(1)\right],\\
I_{31}&=&-\Gamma(1+z_1)\Gamma(-z_1)\left[\psi(1+z_1)-\psi(1)\right]\left(\psi(1-z_1)-\psi(1)\right),\\
I_{32}&=&-\Gamma(2+z_1)\Gamma(-1-z_1)\left[\psi(2+z_1)-\psi(1)-1\right]\left(\psi(1-z_1)-\psi(1)\right).
\end{eqnarray*}
Summing up one finds the result for the integral over $z_4$
\begin{equation}
-2\Gamma(1+z_1)\Gamma(-z_1)\frac{z_1}{1+z_1}\left(\psi(1+z_1)-\psi(1)\right).
\end{equation}
Eventually, one gets the remaining integral over $z_1$
\begin{eqnarray}
DBox_6(s,t)&=&
\left(\frac{\pi^3}{(2\pi)^6}\right)^2\frac{2}{s}\int_{-i\infty}^{i\infty}
\frac{dz_1}{2\pi i} \ x^{z_1}
\left[\Gamma(1+z_1)\Gamma(-z_1)\right]^3\left[\psi(1+z_1)-\psi(1)\right]\nonumber\\
&\times&\left(1-\frac{1}{1+z_1}\right),
\end{eqnarray}
which can be calculated taking the residues at $z_1=0,1,...$ and
evaluating the sum.  The last step  can be performed with the help
of the formulae from \cite{Smirnov Book}, Ch.C. The result is
\begin{equation}
DBox_6(s,t)=
\left(\frac{\pi^3}{(2\pi)^6}\right)^2\left(\frac{b_4(x)}{t}+\frac{b_3(x)}{s+t}\right),
\end{equation}
where the functions $b_i(x)$ are given above and coincide with the
ones obtained in \cite{Tausk} by using differential equations method.

\section{Appendix B}

\begin{figure}[ht]
 \begin{center}
  \epsfxsize=6cm
 \epsffile{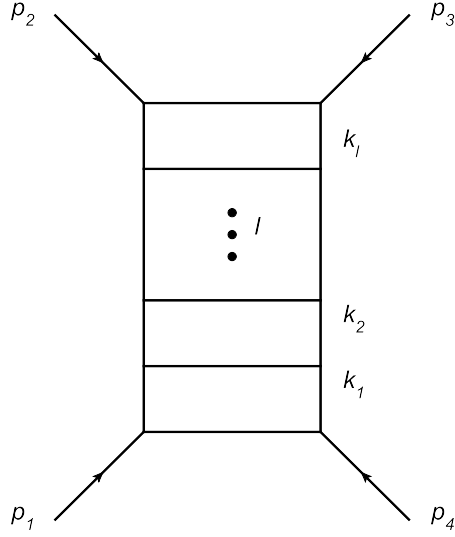}
 \end{center}\vspace{-0.2cm}
 \caption{The Box type scalar integral with l-rungs.}\label{Box_Ladder_pic}
 \end{figure}

Consider the $D=6$ box type scalar integral with l-rungs shown in
Fig.\ref{Box_Ladder_pic}. It is UV/IR finite in all orders of PT. We
are interested in its asymptotics in the Regge limit when $s\to
+\infty$, $t<0$ and fixed.
In what follows we use the evaluation method suggested by E.Kuraev.

First consider the one loop box. It is given by the integral:
\begin{eqnarray}
  B_{l=1}(s,t)=\frac 1i\int \frac{d^6k}{(2\pi)^6}
  \frac{1}{k^2(k-p_2)^2(k+p_1)^2(k+p_1+p_4)^2}.
\end{eqnarray}
Using the  Sudakov variables to parametrize the loop momentum
\begin{eqnarray}
 k= \alpha p_2+\beta p_1 +k_{\perp},
\end{eqnarray}
one gets
\begin{eqnarray}
  d^6k=\frac{s}{2}~d\alpha~d\beta~d^4k_{\perp},
  ~d^4k_{\perp}=k_{\perp}^2~dk_{\perp}^2~d\Omega_4,
\end{eqnarray}
where as usual
\begin{eqnarray}
 s=(p_1+p_2)^2,~t=(p_1+p_4)^2,~s>0,t<0.
\end{eqnarray}
In the limit of $s\gg 1$  $(k+p_1)^2$ and $(k-p_2)^2$ can be estimated as
\begin{eqnarray}
  (k+p_1)^2\simeq s\alpha,~(k-p_2)^2\simeq -s\beta,
\end{eqnarray}
so we can rewrite $B_{l=1}$ as (hereafter we will omit common
$(-\pi^3/(2\pi)^6)^l$ factor)
\begin{eqnarray}
  B_{l=1}(s,t) \simeq \frac{1}{s}\int_{t/s}^1\frac{d\alpha~d\beta}{\alpha\beta}
  \int\frac{d^4k_{\perp}}{k^2(k+p_1+p_4)^2}.
\end{eqnarray}
The  leading asymptotics of the bubble type integral  can be estimated as
\begin{eqnarray}
  \int\frac{d^4k_{\perp}}{k^2(k+p_1+p_4)^2}\simeq \theta(s \alpha \beta -t),
\end{eqnarray}
So for the box integral one gets (remind that $x=s/t$, $L(x)\doteq\log(x)$)
\begin{eqnarray}
  B_{l=1}(s,t) &\simeq&
  \frac{1}{s}\int_{t/s}^1\frac{d\alpha~d\beta}{\alpha\beta}
  \theta(s\alpha \beta -t)=\int_0^{L(x)} da \int_0^{L(x)} db
  ~\theta(a+b-t)\nonumber\\
  &=& \frac{L^2(x)}{s}\int_{0}^1 da~db~\theta(a+b-1)=
  \frac{1}{2}\frac{L^2(x)}{s},
\end{eqnarray}
which is consistent with the explicit result (\ref{b2}).

For the double box $B_{l=2}(s,t)$ using the same approximations
we obtain ($\alpha_i,\beta_i$ correspond to the $d^6k_i$ loop momenta):
\begin{eqnarray}
  B_{l=2}(s,t)&\simeq& \frac{1}{s}\int_{t/s}^1
  \frac{d\alpha_1d\beta_2}{\alpha_1\beta_2}
  \frac{d\alpha_2 d\beta_1}{(\alpha_1-\alpha_2)(\beta_1-\beta_2)}
  \theta(s\alpha_1\beta_1-t)  \theta(s\alpha_2\beta_2-t)\nonumber\\
  &\simeq&\frac{L^4(x)}{s^2}\int^1_0\prod_{i=1}^2da_i db_i
  \theta(a_1+b_1-1)\theta(a_2+b_2-1)\theta(a_1-a_2)\theta(b_2-b_1)
  \nonumber\\
  &=&\frac{1}{12}\frac{L^4(x)}{s},
\end{eqnarray}
which once again is consistent with the explicit result (\ref{DB}).

For the l-rung box integral one can get along the same lines
\begin{eqnarray}
  B_{l}(s,t)\simeq \frac{L^{2l}(x)}{s}\mathcal{I}_l,~~~~l\geq2,
\end{eqnarray}
with
\begin{eqnarray}
  \mathcal{I}_l=\int^1_0\prod_{i=1}^lda_i db_i
  \prod_{k=1}^l\theta(a_k+b_k-1)
  \prod_{p=1}^{l-1}\theta(a_p-a_{p+1})
  \prod_{m=1}^{l-1}\theta(b_{m+1}-b_m).
\end{eqnarray}
The easiest way to treat this integral is to  evaluate it numerically for
several values of $l$. The result coincides with the analytical formula
\begin{eqnarray}
  \mathcal{I}_l=\frac{1}{l!(l+1)!}.
\end{eqnarray}
So, finally, we have the following result for the leading
logarithmic asymptotics of the $D=6$ l-rung box function:
\begin{eqnarray}
 B_l(s,t)\simeq\frac{1}{l!(l+1)!}\frac{L^{2l}(x)}{s}.
\end{eqnarray}

\section*{Acknowledgements}
We are grateful to A.Davydychev and V.Smirnov for consultations
concerning the integration technique and to A.Pikelner for his help
in numerical verification of the results. Discussions of the Regge
limit and methods to evaluate the asymptotics of the ladder diagrams
with L.Lipatov and E.Kuraev are highly appreciated. L.V. would like
to thank A.Gorsky and I.Samsonov for useful comments and stimulating discussions and
S.Ogarkov for interesting discussions. Financial support from the
RFBR grant \# 11-02-01177 and the Ministry of
Education and Science of the Russian Federation grant \# 3802.2012.2
is kindly acknowledged. D.V. is grateful to Dynasty Foundation for support.

\end{document}